\pdfoutput=1
\documentclass[a4paper,american,citeautoscript,floatfix,pdftex,superscriptaddress,twoside,%
aps,pra,%
reprint
]{revtex4-2}%
\usepackage{amsfonts,amsmath,amssymb}
\usepackage[T1]{fontenc}
\usepackage{graphicx}%
\usepackage[utf8]{inputenc}
\usepackage[ams,todos]{CMImacros}
\usepackage{hypernat}

\graphicspath{{./figures/}{./fig/}} 
\hypersetup{citebordercolor=yellow,linkbordercolor=red,urlbordercolor=blue} %

\newcommand{\cfeldesy}{\affiliation{Center for Free-Electron Laser Science, Deutsches
      Elektronen-Synchrotron DESY, Notkestraße 85, 22607 Hamburg, Germany}}%
\newcommand{\uhhcui}{\affiliation{Center for Ultrafast Imaging, Universität Hamburg, Luruper
      Chaussee 149, 22761 Hamburg, Germany}}%
\newcommand{\uhhphys}{\affiliation{Department of Physics, Universität Hamburg, Luruper Chaussee 149,
      22761 Hamburg, Germany}}%
\newcommand{\mpsd}{\affiliation{Max Planck Institute for the Structure and Dynamics of Matter,
      Luruper Chaussee 149, 22761 Hamburg, Germany}}%
\newcommand{\jkemail}{\email[]{jochen.kuepper@cfel.de}}%
\newcommand{\cmiweb}{\homepage{https://www.controlled-molecule-imaging.org}}%

\newcommand*{\suppinf}{Supplementary Information\xspace}%

\begin{document}
\title{Optimizing the geometry of aerodynamic lens injectors for single-particle coherent
   diffractive imaging of gold nanoparticles}%
\author{Lena Worbs}\cfeldesy\uhhphys%
\author{Nils Roth}\cfeldesy\uhhphys%
\author{Jannik Lübke}\cfeldesy\uhhphys\uhhcui%
\author{Armando D.~Estillore}\cfeldesy%
\author{P.~Lourdu Xavier}\cfeldesy\mpsd%
\author{Amit K.~Samanta}\cfeldesy%
\author{Jochen Küpper}\jkemail\cmiweb\cfeldesy\uhhphys\uhhcui%
\begin{abstract}
   Single-particle x-ray diffractive imaging (SPI) of small (bio-)nanoparticles (NPs) requires
   optimized injectors to collect sufficient diffraction patterns to reconstruct the NP structure
   with high resolution. Typically, aerodynamic-lens-stack injectors are used for single NP
   injection. However, current injectors were developed for larger NPs ($\gg\!100$~nm) and their
   ability to generate high-density NP beams suffers with decreasing NP size. Here, an
   aerodynamic-lens-stack injector with variable geometry and the geometry-optimization procedure
   are presented. The optimization for 50~nm gold NP (AuNP) injection using a numerical simulation
   infrastructure capable of calculating the carrier gas flow and the particle trajectories through
   the injector is introduced. The simulations are experimentally validated using spherical AuNPs
   and sucrose NPs. In addition, the optimized injector is compared to the standard-installation
   ``Uppsala-injector'' for AuNPs and results for these heavy particles show a shift in the particle-beam 
   focus position rather than a change in beam size, which results in a lower gas background
   for the optimized injector. Optimized aerodynamic-lens stack injectors will allow to increase NP
   beam density, reduce the gas background, discover the limits of current injectors, and contribute
   to structure determination of small NPs using SPI.
\end{abstract}
\maketitle

\section{Introduction}
Simulations predicted the possibility of deriving high-resolution structures of biological
macromolecules using x-ray free-electron lasers (XFELs)~\cite{Neutze:Nature406:752}. The ultra-short
and extremely bright pulses of coherent x-rays provided by free-electron lasers (FELs) can outrun
radiation damage processes before the particle has time to structurally respond and eventually be
destroyed by the deposited energy~\cite{Chapman:NatPhys2:839}. Thus, the single-particle diffractive
imaging (SPI) method at XFELs can be used to elucidate the structure of biological
molecules~\cite{Bogan:NanoLett8:310, Aquila:StructDyn2:041701} without the need of highly ordered
crystalline sample. SPI allows to retrieve the three-dimensional (3D) structure of biomolecules by
reconstruction from a large number of two dimensional diffraction patterns, assembled into a 3D
diffraction volume, requiring a high probability of an x-ray pulse interacting with an injected
particle~\cite{Ayyer:Optica8:15}. High-density particle-beams with ideally one particle per pulse
and focus volume are generated to use both, x-rays and sample, efficiently. However, for the atomic
resolution, $\ordsim100$~pm, reconstruction of a protein, $10^5$ to $10^6$ diffraction patterns need
to be collected~\cite{Poudyal:StrDyn7:024102}.

Delivery of high-density single-particle-beams was demonstrated using aerodynamic-lens stacks (ALS)
to generate focused beams of aerosolized particles from ambient conditions into
vacuum~\cite{Bogan:NanoLett8:310, Ayyer:Optica8:15}. An ALS contains sets of thin apertures to
manipulate the particles’ lateral spatial distribution before it exits through the last aperture
into vacuum. Aerodynamic lenses enable successive contractions of a flowing particle-beam and
provide focusing to high particle-densities for wide range of particle sizes~\cite{Bogan:AST44:i,
   Benner:JAerosolSci39:917}. Before adaption for SPI, they were mainly used in aerosol mass
spectrometry to ensure a high transmission for a large particle size
range~\cite{Canagaratna:MSR26:185}. A widely used ALS for SPI, readily available at many XFELs, is
the so-called ``Uppsala injector'' (TSI model AFL100), which can deliver collimated or focused beams for a range of
particle sizes, \eg, 0.1--300~\um~\cite{Hantke:NatPhoton8:943}. It was successfully used in various
experiments at XFEL facilities~\cite{Ho:NatComm:112020, Bielecki:SciAdv5:eaav8801,
   Hantke:NatPhoton8:943, Seibert:Nature470:78}. A recent experiment we performed at EuXFEL showed
the successful collection of more than 10 million diffraction patterns from single gold
nanoparticles using this injector and shows the opportunities provided by careful sample preparation
and injection~\cite{Ayyer:Optica8:15}.

However, currently sample injection and beam formation is the bottleneck of collecting large data
sets of small bio-particle diffraction patterns and injection schemes have to be modified
accordingly~\cite{Bielecki:StrDyn:7040901}. The geometry of the ``Uppsala injector'' is fixed by
design and the only tunable parameter to change the particle-beam focus size and position is the
inlet pressure before the ALS~\cite{Hantke:IUCr5:673}. To circumvent the increase of inlet pressure
to generate a smaller particle-beam focus and thus an increase of pressure in the experimental
chamber, we designed and used a new particle injector with variable geometry, as shown in
\autoref{fig:als_geometry}, \ie, the inner tube diameter and the aperture diameter can be
changed~\cite{Roth:JAS124:17, Roth:flash-nano-injector-polystyrene-beams:inprep} to produce the
highest particle-beam density for a given particle size. In addition, the speed of the particles is
important in SPI experiments. It should be as slow as possible to increase the particle-beam density
and thus hitrate, but with increasing repetition rates at XFEL facilities the particle speed has to
be sufficiently fast to avoid interaction of NPs with two x-ray pulses. For the full repetition rate
of 4.5~MHz at EuXFEL~\cite{Decking:NatPhot14:391} and an x-ray focus size of 2~\um, the particle
speed has to exceed 10~m/s to enter the interaction region without interacting with the previous
pulse, which can damage or scatter off the sample already.

Here, we present the geometry optimization for aerosolized spherical gold nanoparticles (AuNPs) of
50~nm diameter at typical inlet conditions for SPI experiments. The size is chosen as an intermediate 
step from large NPs and viruses ($>100$~nm) towards single proteins ($<10$~nm). The numerical simulation
infrastructure used is presented elsewhere~\cite{Roth:JAS124:17, Welker:CMInject:inprep}. To
validate our simulation results, we compare them with experimental data for both AuNPs and sucrose
spheres. AuNPs, when synthesized and prepared well, show a narrow size distribution similar to
bio-particles and are therefore good benchmark samples for sizing and focusing experiments. AuNPs
have a high scattering power resulting in good detection efficiencies both in x-ray scattering and
for in-laboratory detection methods~\cite{Awel:OptExp24:6507, Worbs:OptExp27:36580}. Furthermore,
AuNPs exhibit distinct physical and chemical properties with potential applications ranging from
quantum electronics to biomedicine and potential drug delivery
systems~\cite{Schmid:ChemComm:B411696H, Dykman:CSR41:2256}. Sucrose particles are often used at XFEL
facilities for alignment in commissioning and startup experiments~\cite{Ho:NatComm:112020,
   Bielecki:SciAdv5:eaav8801}, as the number density of the generated sucrose spheres is high and
the particle-beam can be observed easily while aligning the injector to the x-ray beam. Most
importantly, the mass density of sucrose NPs, and thus their
focusing behavior, is comparable to biological matter, rendering them a good prototypical benchmark
system for bio-nanoparticles.

\section{Methods}
\subsection{Geometry Optimization}
\begin{figure}
   \includegraphics{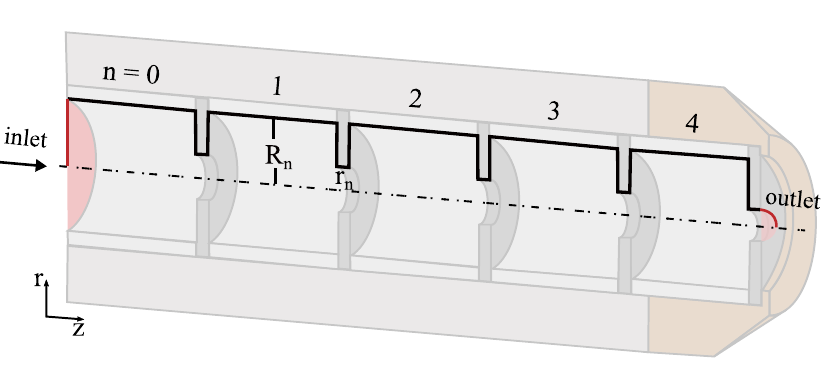}%
   \caption{Schematic of the ALS geometry, which is cylindrically symmetric about the dashed line.
      Carrier gas flows from left to right. The black solid line is the 2D geometry used for the
      simulations, consisting of five aperture and tube pieces. The inner radius ($R_n$) of the tube
      as well as the lens aperture radius ($r_n$) can be changed
      individually~\cite{Roth:flash-nano-injector-polystyrene-beams:inprep}; see text for details.}
   \label{fig:als_geometry}
\end{figure}
Simulations of the ALS were performed as follows: First, we calculated the flow field of the carrier gas inside a
given 2D cylindrically symmetric geometry using a finite-element solver for the Navier-Stokes
equations~\cite{Comsol:Multiphysics:5.5}. The flow field was calculated within the ALS geometry and
extended after the exit with a quarter-circle with the radius of the last aperture serving as
gas-expansion region of the vacuum chamber as shown in \autoref{fig:als_geometry}. The carrier gas
was assumed to be nitrogen, as the particles were aerosolized using electrospray ionization (ESI), where the
used gas mixture consists of $\approx$ 90~\% nitrogen and $\approx$ 10~\% CO$_2$. As boundary
conditions for the flow field we used mass-flow conservation of 13~mg/min as inlet condition and a
pressure of 10$^{-4}$~mbar at the end of the flow field along the semi-circle. Additional flow field
calculations were performed using the inlet pressure as a boundary condition. Second, the
trajectories of 100,000 particles for a given flow field were calculated with a homebuilt Python
particle-tracing code~\cite{Roth:JAS124:17}. Particles were introduced into the flow field with a
uniform radial distribution covering the diameter of the first tube piece. We assumed the particles'
velocity to be equal to the flow field values. We simulated trajectories of 50~nm diameter spherical
particles with a density of 19.32 g/cm$^3$, corresponding to the bulk density of gold. Transmitted
particles were propagated further with their terminal speeds at the border of the
flow field. Then, we determined the width of the resulting particle-beam depending on the distance
from the ALS exit, \ie, the last aperture. Beam widths $d_{70}$ were
specified as the diameter where 70~\% of the particles were in; $d_{70}$ is a useful and robust
metric as it is independent of the actual beam shape. Nevertheless, outside of the ALS all simulated
particle-beams showed a peak-like radial distribution with the maximum of the
particle density in the center ($r=0$~mm).

The ALS consists of $n=5$ aperture/tube
pieces stacked onto another, see \autoref{fig:als_geometry}. The lens aperture radius
$r_n$ and the inner tube radius $R_n$ can easily be adjusted. In our ALS, the aperture radius can 
be chosen from 0.75~mm to
5~mm in 0.25~mm steps. The lens apertures are interchangeable. The inner tube diameter could be 
chosen from parts with radius $2,3,4,5,6,7.5,8,10$~mm, which were available in stock; in principle, 
any size is possible. The inner diameter of the tubes is adjusted by adding an additional tube into the 
standard 10~mm diameter pieces.

As the variety corresponds to more than $7\times10^{10}$ combinations we approached the optimization
as follows: Our optimization procedure was performed iteratively from the exit to the entrance of the ALS, 
as the last aperture radius ($r_4$) largely determines the focus position and size of the particle
beam. A larger $r_4$ aperture moves the particle-beam focus
further outside or creates a collimated particle-beam and a smaller $r_4$ creates a particle-beam
focus closer to the aperture or an even diverging particle-beam. We started the optimization in the
last piece of the ALS, $r_4$ and $R_4$. The particles were introduced into the flow field with a
uniform radial distribution covering $r_{\text{initital}}=0.02$~mm,
mimicking that the lenses before already prefocused the particle-beam. The
initial particle velocity was set equal to the flow fields speed. The best $r_4$ and $R_4$
combination fulfills the following condition: The transmission was $\larger90$~\%, the focus was at
$z>4$~mm to suppress background scattering from the housing of the ALS, and it resulted in the
smallest beam diameter. With this optimized $r_4$ and $R_4$ combination we then optimized $r_3$ and
$R_3$, and this was subsequently iterated for all lenses with increasing initial radial distribution
of the particles, \ie, 0.02~mm for piece 4 and 3, 0.5~mm for piece 2, 1~mm for piece 1, and the
whole radius of the lens filled before the first aperture. This optimization
procedure reduces the efforts to 160 combinations per lens and $\smaller1000$ overall.

\subsection{Experimental Setup}
\begin{figure}
   \includegraphics{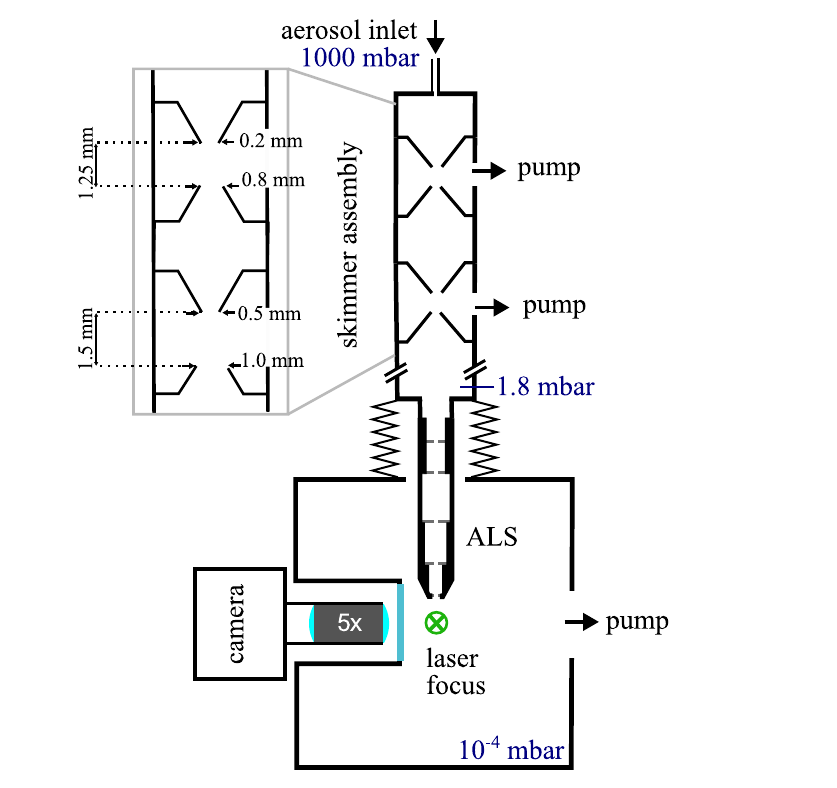}%
   \caption{Schematic of the experimental setup for the characterization of nanoparticle-beams. The
      aerosol passed a skimmer assembly to remove most of the carrier gas and the particles were
      focused using an ALS and entered the main vacuum chamber, where the particle-beam was crossed
      by a laser beam. The light scattered off the particles was collected using a camera-based
      microscope system~\cite{Awel:OptExp24:6507, Worbs:OptExp27:36580}.}
   \label{fig:setup}
\end{figure}
We measured the particle-beam evolution of AuNPs from the optimized ALS geometry. The schematic of
our experimental setup is shown in \autoref{fig:setup}. It consists of four main parts: an
aerosolization chamber, a differentially pumped transport tube, the ALS system for particle-beam
formation, and the detection region for visualization of the particle-beam. To generate isolated test 
particles from the liquid sample, we injected spherical AuNPs with a diameter of $(27\pm2.25)$~nm in
5~mM ammonium acetate (AmAc) with a concentration of $10^{11}$~particles/ml and a 2~\% sucrose
solution in 20~mM AmAc using a commercial electrospray (TSI Advanced Electrospray 3482). The
aerosolized nanoparticles passed through a differentially pumped skimmer assembly for pressure
reduction. The particles were focused into the detection chamber using the ALS. The pressure above
the entrance of the the ALS was 1.8~mbar (Pfeiffer Vacuum CMR 361). In the main chamber, the
pressure was kept at $2.5\times10^{-4}$~mbar. The ALS is mounted on a motorized $xyz$-manipulator to
perform height scans and measure the particle-beam evolution.

Particles were detected using a side-view illumination scheme~\cite{Awel:OptExp24:6507}. A Nd:YAG
laser (Innolas SpitLight, 532~nm, pulse duration 11.5~ns, pulse energy up to
240~mJ at 532~nm, 20~Hz repetition rate) was focused into the center of the vacuum chamber
intersecting the particle-beam. The light scattered off the particles was collected using a
camera-based microscope system~\cite{Awel:OptExp24:6507, Worbs:OptExp27:36580} consisting of a long
working-distance objective (Edmund Optics, $5\times$ magnification, numerical aperture $N_a=0.14$,
working distance $d=34$~mm, depth of field $14~\um$) and a high-efficiency sCMOS camera
(Photometrics PrimeB95, quantum efficiency 0.95 at 532~nm, $1200\times1200$ pixels). This microscope
yields a nominal resolution of 0.54~pixel/\um. Images were collected with a 1~ms exposure time
synchronized to the laser at 20~Hz such that every frame covered one laser pulse. For every distance
of the ALS and the laser, we recorded 10000 images for the AuNP sample and 2000 images for the
sucrose sample. We determined the positions of the particles by analyzing the images using a
centroiding algorithm based on Hessian blob-finding~\cite{Marsh:scirep8:978}. The particles'
positions were converted into a 2D histogram, see \suppinf for details. The width of the particle
beam is determined from the projection of the particle-beam onto the laser propagation axis. The
beam diameter is shown as $d_{70}$.

\section{Results}
\subsection{Optimization and Simulation Results}%
\begin{figure*}
   \includegraphics{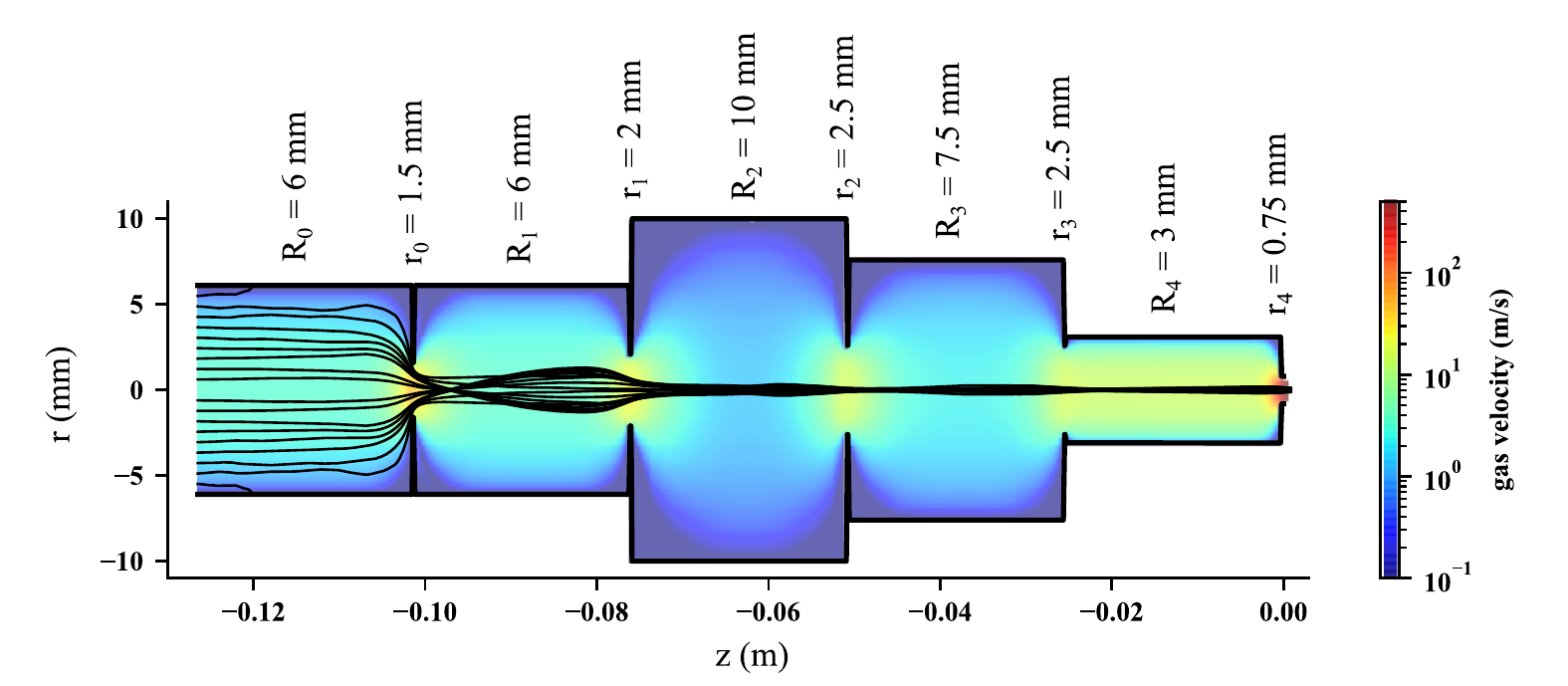}
   \caption{Optimized ALS geometry. Tube and aperture radii are specified above the device and the
      corresponding nitrogen-gas flow-field for the injection of 50~nm AuNPs at 13~mg/min mass flow
      is depicted in false color. Representative (calculated) particle trajectories are shown by black lines,
      with gas and particle flow direction from left to right. A clear focusing effect of the
      different parts of the ALS can be observed through the radial narrowing of the set of particle
      trajectories.}
   \label{fig:opt_geometry}
\end{figure*}
The optimization process resulted in one final geometry which produced a particle beam to our 
specifications. The resulting optimized-lens-stack geometry is shown in \autoref{fig:opt_geometry}. 
From entrance to exit, the lens tube radius and the aperture radius are first increasing, then decreasing. 
The smallest lens tube radius and aperture radius are located at the last lens piece. Values are given 
next to the geometry. 
The velocity-flow field for 13~mg/min mass flow and particle trajectories for 50~nm AuNPs at different 
inlet positions (black solid lines) are shown in \autoref{fig:opt_geometry}, demonstrating a clear focusing 
effect of the ALS.

\begin{figure}
   \includegraphics{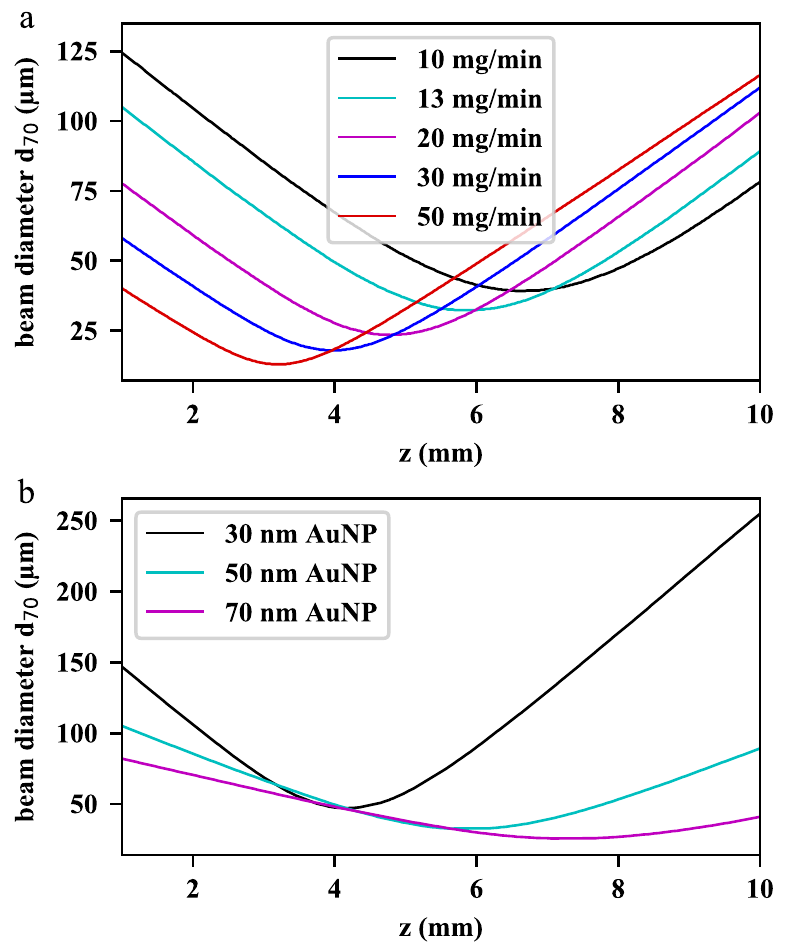}
   \caption{(a) Particle-beam evolution curves of the optimized injector for 50~nm AuNPs at
      different gas-mass flows. The width of the particle-beam was determined as $d_{70}$. With
      increasing mass flow and thus pressure before the ALS, the particle-beam focus becomes harder,
      \ie, it moves closer to the ALS exit and gets smaller. (b) Particle-beam evolution curve of
      the optimized injector for different AuNP sizes at 13~mg/min mass flow. With increasing
      particle size, the particle-beam focus decreases and moves further away from the ALS. The
      convergence increases with decreasing particle size.}
   \label{fig:focusing}
\end{figure}
For this ALS geometry, the 50~nm AuNP beam focused at a distance of 5.8~mm from the ALS exit with a
particle-beam width of of 33~\um ($d_{70}$). The particle-beam evolution for 13~mg/min mass flow is
shown in \autoref[a]{fig:focusing} as the cyan curve. The particle-beam evolution is shown as the
beam width ($d_{70}$) depending on the distance $z$ from the ALS exit. We simulated the focusing
behavior for different mass flow conditions between 10 and 50~mg/min. With increasing mass flow, the
focus shifted closer to the exit of the ALS and the focus size decreased. At 50~mg/min mass flow,
the particle-beam focus size decreased to 13~\um at a distance of 3.2~mm. Similar behavior has been
shown experimentally for the ``Uppsala-injector''~\cite{Hantke:IUCr5:673}. Therefore, working at
higher mass flow is desired, but it will increase the amount of gas introduced into the interaction
chamber and result in a higher pressure and thus a higher gas-scattering background in diffractive
imaging experiments.

At 13~mg/min mass flow the AuNPs exiting from the optimized ALS had a mean velocity of 29~m/s. Mean
velocities and beam diameter values for different flow conditions are given in the \suppinf. The
behavior of the ALS optimized for 50~nm AuNPs was compared to smaller and larger diameters of the
AuNPs as shown in \autoref[b]{fig:focusing}. For smaller AuNPs the focus moved closer to the ALS and
was larger, whereas bigger particles were focused further away and showed a smaller focus size. An
interesting feature observed was the change of the convergence depending on the particle size:
The smaller the particles were, the larger the convergence became. A precise positioning for small 
particles becomes necessary to meet the particle-beam focus. This change of the convergence is due to the
larger momentum of larger particles interacting with the gas flow field.

\subsection{Experimental Results}
\begin{figure}
   \includegraphics{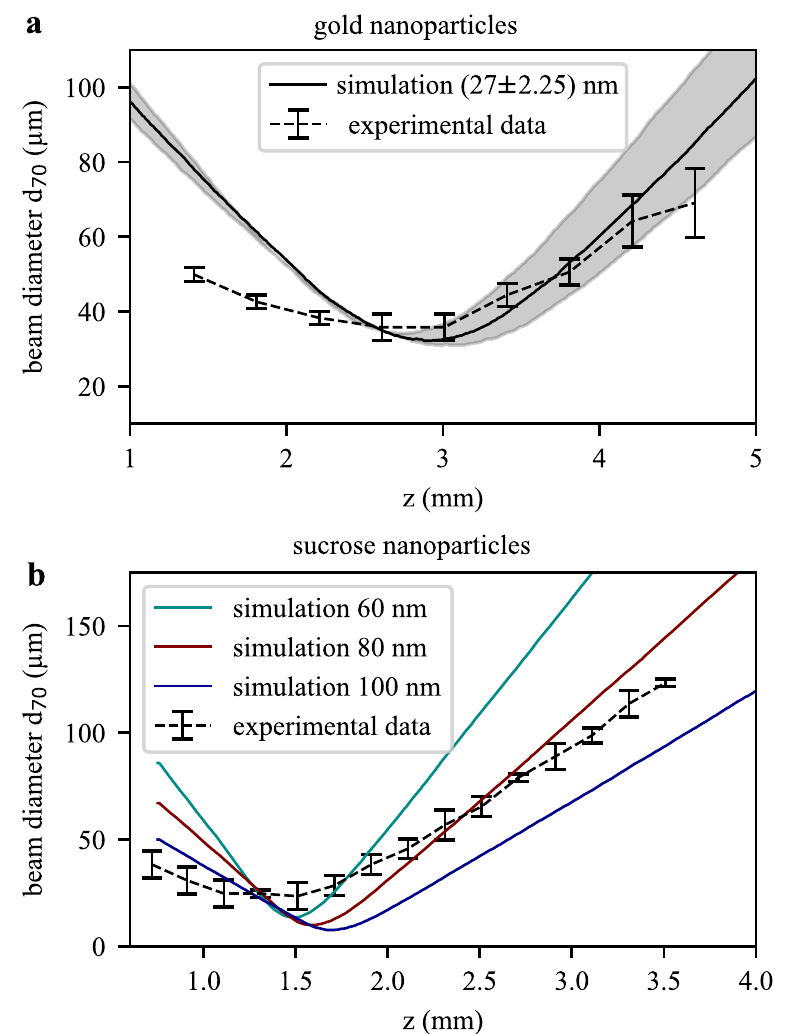}%
   \caption{(a) Experimental particle-beam size evolution for (27$\pm$2.25)~nm AuNPs (black dashed
      line). Simulated beam evolution is shown for 27~nm (black solid line) with the spread of the
      beam diameter due to the size distribution of $\pm$2.25~nm (grey area). (b) Experimental
      particle-beam size evolution for sucrose spheres (black). The experimental data agrees
      reasonably well with a simulated particle size of 80~nm (dark red). }
   \label{fig:exp}
\end{figure}
We measured particle-beam evolution curves of AuNPs and sucrose particles. The AuNP data with
standard errors is shown in \autoref[a]{fig:exp} as beam diameter ($d_{70}$) depending on the
distance from the injector exit, along with simulations for the particle size of $(27\pm2.25)$~nm using 
the experimentally measured inlet pressure of 1.8~mbar above
the ALS. The experimental and simulated particle-beam diameters agree well, especially at and after
the focus of the particle-beam. Some deviations are observed before the focus, where the simulation
overestimates the beam diameter, \eg, by a factor of $\ordsim1.5$ at $z=1.4$~mm. However, the most
relevant parameters for SPI experiments, the focus position and focus size, are in excellent
agreement between experiment and simulation. The same experiment is repeated for a 2~\% sucrose
solution to generate spherical sucrose particles in the electrospray process with a broad size
distribution around 80~nm, shown in the \suppinf. The sucrose-particle-beam evolution is shown in
\autoref[b]{fig:exp} with standard errors (black) and compared to simulations for different sizes of
sucrose spheres ($\rho=1.59$~g/cm$^3$). Overall, the experimental data is described well by the
simulation for 80~nm sucrose spheres. Similar to the AuNP data, the simulation agrees well with our
data after the focus, although before the focus the simulation deviates by a factor of $\ordsim1.6$
at $z=0.8$~mm. This mismatch is partly due to the broad experimental size distribution, \ie, the
experimental data does not correspond to a single particle-size simulation.

\subsection{``Uppsala-injector'' simulation} %
The ``Uppsala-injector'' (AFL100) was
introduced before and used in various experiments at XFELs~\cite{Ho:NatComm:112020,
   Bielecki:SciAdv5:eaav8801, Hantke:NatPhoton8:943, Seibert:Nature470:78}. We simulated its
focusing of 50~nm AuNPs and compared it to our optimized ALS.

\begin{figure}
   \includegraphics{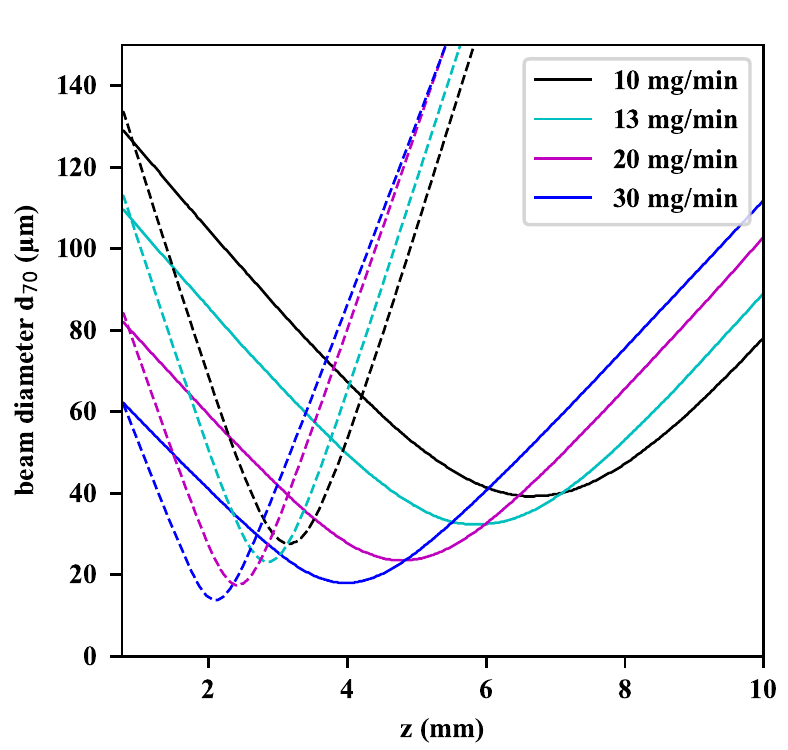}
   \caption{Simulated particle-beam evolution curves of 50~nm AuNPs exiting from the
      ``Uppsala-injector'' (dashed lines) for different mass flow conditions in comparison to the
      corresponding focusing curves from our optimized injector (solid lines).}
   \label{fig:Uppsala}
\end{figure}
The beam evolution curves for 50~nm AuNPs at different mass flow conditions for this injector are
shown in \autoref{fig:Uppsala} (dashed lines), along with the focusing curves for our optimized
injector (solid lines). The simulated transmission of 50~nm AuNPs through both injectors was above
90~\%, \ie, at 13~mg/min mass flow the transmission was 91.9~\% for the AFL100 and 93.4~\% for our
optimized injector. At the same mass flow the AFL100 showed slightly smaller mean velocities of the
exiting particles than our optimized injector. As an example, at 13~mg/min mass flow, 50~nm AuNPs
exiting the ``Uppsala-injector'' showed a mean velocity of 27~m/s. In comparison, particles in our
injector reached a velocity of 29~m/s. A detailed list of velocities depending on mass flow is shown
in the \suppinf.

\section{Discussion}%
Our simulations show that the focus size is comparable for both injectors, but the difference is in
the focus position and the convergence. Our injector focuses the particles further downstream and
the focusing is not as hard as for the AFL100. Generating a focused particle-beam further away from
the injector exit has the advantage of a lower background from the nitrogen and CO$_2$ gas. Light
gas diverges fast from the exit of the ALS into the vacuum chamber.

The focus position of the AFL100 is closer to the injector exit and can cause problems when using
smaller particles and particles with lower sample density, such as bio-particles: The smaller and
lighter the particles are the closer the focus position. As an example, we performed simulations for
10~nm AuNPs at 13~mg/min nitrogen mass flow for the AFL100 showing the focus position moves very
close to the ALS exit, below $z=1.5$~mm; the transmission is reduced to 59~\%. Our 50~nm-optimized
injector still shows a transmission of 79~\% for those particles and a focus position above $z=2$~mm; see
\suppinf for details. The same behavior holds for the particle density: The lower the particle
density (biomolecules), the closer the particle-beam focus becomes. For isolated proteins, it is
almost impossible to focus the particle-beam with these injectors. In this case, an appropriate
geometry optimization could result in a particle-beam focus further away from the injector exit.

\section{Conclusion} %
We presented an optimization procedure of an ALS for 50~nm AuNPs using our previously developed
computer-simulation framework for ALS injectors~\cite{Roth:JAS124:17, Welker:CMInject:inprep}
including the results of this optimization. We experimentally benchmarked the optimized geometry for
beams of spherical gold and sucrose nanoparticles. Both particle-beam-evolution curves are in good
agreement with the simulations. This validates our simulation framework, which can be used to get
further insight into the fluid-dynamics focusing process and to develop optimized particle injectors
for different sizes and materials, as well as for different experimental conditions, such as inlet
pressure and gas type.

We compared our optimized injector to the widely used AFL100 ``Uppsala-injector'' for 50~nm AuNPs.
Both injectors create a focused particle-beam for different inlet mass flow conditions, and the main
difference is observed in the particle-beam focus position, which for our optimized injector is
further downstream, which reduces the carrier gas background at the focus and will be greatly
beneficial for x-ray diffractive imaging, especially of small bio-particles that exhibit only small
scattering signals.

Our variable injector geometry allows us to vary the particle-beam focus independent from the inlet
pressure by varying the geometry and thus keeping the pressure after the injector, \ie, in the x-ray
interaction region, constant. Generating high-quality particle-beams of nanoparticles does not only
allow for structure determination by an increased number of collected diffraction patterns, but in
addition it open the field of time-resolved imaging of nanoparticle dynamics in future
pump-probe-type experiments at XFELs.

\bigskip

\appendix
\section*{Acknowledgement}
We thank Simon Welker for helpful discussions on improvements of the particle trajectory
calculations. This work has been supported by the European Research Council under the European
Union's Seventh Framework Program (FP7/2007-2013) through the Consolidator Grant COMOTION (614507)
and the Cluster of Excellence ``Advanced Imaging of Matter'' (AIM, EXC~2056, ID~390715994) of the
Deutsche Forschungsgemeinschaft (DFG). Parts of this research were supported by the Maxwell
computational resources operated at DESY. P.~Lourdu Xavier acknowledges a fellowship from the
Joachim Herz Stiftung.

\bibliography{string,cmi}

\onecolumngrid
\listofnotes
\end{document}